# CNN-based fully automatic wrist cartilage volume quantification in MR Images

Nikita Vladimirov[1], Ekaterina Brui[1], Anatoliy Levchuk[1,2], Vladimir Fokin[1,2], Aleksandr Efimtcev[1,2], David Bendahan[3]

[1]School of Physics and Engineering, ITMO University, Saint-Petersburg, Russia

[2]Department of Radiology, Federal Almazov North-West Medical Research Center, Saint-Petersburg, Russia

[3]Centre de Résonance Magnétique Biologique et Médicale, Aix-Marseille Universite, CNRS, Marseille, France

**Correspondence**: Ekaterina Brui, 191002, St Petersburg, Russia, Lomonosova st. 9. Email: e.brui@metalab.ifmo.ru

## Abstract

### Background and objective

Detection of cartilage loss is crucial for the diagnosis of osteo- and rheumatoid arthritis. A large number of automatic segmentation tools have been reported so far for cartilage assessment in magnetic resonance images of large joints and several MRI biomarkers of cartilage loss have been proposed. As compared to knee or hip, wrist cartilage has a more complex structure so that automatic tools developed for large joints are not expected to be operational for wrist cartilage segmentation. In that respect, a fully automatic wrist cartilage segmentation method would be of high clinical interest.

### Methods

We assessed the performance of four optimized variants of the U-Net architecture with truncation of its depth and addition of attention layers (U-Net_AL). The corresponding results were compared to those from a patch-based convolutional neural network (CNN) we previously designed. The segmentation quality was assessed on the basis of a comparative analysis with manual segmentation (ground truth) using several morphological (2D DSC, 3D DSC, precision) and a volumetric metrics. The proposed networks were compared using a cross-validation approach on a dataset of 33 3D VIBE images of mostly healthy volunteers. Influence of some image parameters on the segmentation reproducibility was assessed.

### Results

The four networks outperformed the patch-based CNN in terms of segmentation homogeneity and quality. The median 3D DSC value computed with the U-Net_AL (0.817) was significantly larger than the corresponding DSC values computed with the other networks. In addition, the U-Net_AL CNN provided the lowest mean volume error (17%) and the highest Pearson correlation coefficient (0.765) with respect to the ground truth. Of interest, the reproducibility computed from using U-Net_AL was larger than the reproducibility of the manual segmentation. Moreover, the results indicate that the MRI-based wrist cartilage volume is strongly affected by the image resolution.

### Conclusions

U-net convolutional neural network with additional attention layers provides the best wrist cartilage segmentation performance. In order to be used in clinical conditions, the trained network can be fine-tuned on a dataset representing a group of specific patients. The error of cartilage volume measurement should be assessed independently using a non-MRI method.

## 1. Introduction

Multiple morphological metrics computed from magnetic resonance (MR) images have been reported with the aim of assessing the severity of inflammatory and degenerative diseases in articular joints. Quantitative MR-morphometry has been used for linear, surface, and volumetric measurements in bones [1] [2], [3], ligaments [4], meniscus [5] and articular cartilage [6]–[8]. An accurate morphometric approach is expected to allow the detection of dynamic morphological changes related to the disease evolution or in response to a treatment so that the corresponding features could be used as imaging biomarkers [5], [7], [9]. One has to keep in mind that the prerequisite for such a morphometric approach is related to an accurate delineation of the corresponding structures. This task, commonly performed manually, is tedious, time-consuming and suffers from between and within operators variability [6], [10].

Over the last decades, MR morphometry has been used for large joints such as knee, hip and shoulders [11] likely because MR images of the corresponding joints had appropriate spatial resolution, contrast (CNR) and signal-to-noise (SNR) ratios. The accurate delineation of more complex and/or thinner joints from hand, wrist or feet requires images with a resolution that was not technically achievable until recently. Supportive of that, wrist cartilage assessment using MRI was not part of the Rheumatoid Arthritis MRI score until recently [12]. Technical developments such as those in the field of radiofrequency coils have provided the opportunity to obtain images with a higher quality that could be used for a proper assessment of wrist cartilage [13], [14]. The initial segmentation approaches of wrist cartilage were manual with the commonly acknowledged caveats related to duration and reproducibility [6], [15]. More recently, automatic and semi-automatic segmentation methods have been reported for large structures such as bone [3], [16], [17] and very few of them have been reported for wrist cartilage likely as a result of the corresponding complex anatomy. An accurate segmentation method of wrist cartilage would be of high interest especially in rheumatoid arthritis, in which the pathological process has been reported to start in small joints of hand [18] and to affect cartilage rather than bone [19].

The most promising methods for fully automatic segmentation of complex biomedical structures are based on convolutional neural networks (CNNs) [20]–[22]. Several network architectures have shown excellent results for cartilage segmentation from MR images of knee [23] with Dice similarity coefficients up to 0.92 [24], [25], performance for wrist cartilage segmentation has been poorly reported. In a seminal study, Brui *et al*. reported a CNN patch-based approach dedicated to the automatic segmentation of wrist cartilage in 3D MR images [10]. The planar CNN architecture provided an optimized segmentation of centrally-located slices with DSC values up to 0.81 whereas the corresponding segmentation for more lateral slices was poor thereby resulting in a low 3D segmentation accuracy (3D DSC = $0.69 \pm 0.06$). Overall, cartilage volume measurements over the whole 3D space was affected by large errors and other alternatives would be of high interest.

Fully convolution networks (FCNNs) have been largely used in the field of biomedical imaging. U-Net, the most popular architecture [24], [26]–[30], is an autoencoder type network with skip-connections that transfer spatial contextual information from encoder to decoder, so that the contextual information of a whole image is merged with the final features thereby optimizing the spatial relationships recovery at the decoder level [30]. U-Net has been already used for wrist cartilage segmentation and the corresponding results were poor (DSC=0.64) [10] likely because of skipping the step of hyperparameters optimization, which is a crucial step in CNN-based segmentation tasks.

Of interest, MR images of joints often contain structures such as skin and vessels, which display a contrast similar to cartilage so that they could be wrongly identified (false-positive) as cartilage [10]. Attention layers [29], [31], [32] added to a CNN architecture have been developed so that the CNN can learn from relevant regions only.

Such an approach has provided interesting results for the delineation of pancreas [31] and for the segmentation of knee menisci [29]. Such an approach has never been reported for wrist cartilage segmentation.

In the present study, we intended to assess whether wrist cartilage segmentation can be improved with an optimization of U-Net hyperparameters combined to the addition of attention layers. The performance of several CNN architectures was compared to the patch-based CNN approach we initially reported [10]. Performance was assessed on the basis of both conventional segmentation metrics such as DSC and cartilage volume measurements.

## 2. Methods

### 2.1. Datasets and Subjects

Two datasets referred as "small" (*SDS*) and" large" (*LDS*) were used. The SDS was used for an initial comparative analysis regarding segmentation homogeneity within the wrist volume among the tested networks and a previously proposed PB-CNN [10]. This dataset contained MR images obtained from a single 1.5 T MRI scanner and a single imaging sequence (3D VIBE). Given that this homogeneity could be considered as a limitation, we extended the SDS by adding MR images acquired with another scanner (3T) and also images obtained with the same scanner but with a different image resolution (see the details in Supplementary Materials). This enriched dataset was referred as the *LDS*.

The *SDS* consisted of 560 2D slices selected from 20 3D MR images of wrist recorded in 11 subjects (8 healthy and 3 with confirmed osteo- or rheumatoid arthritis) [10]. Images from the *LDS* (1297 2D slices) were those from the *SDS* plus 10 3D MR images recorded in 5 subjects (4 healthy volunteers and one patient with a confirmed RA). Some subjects were scanned twice and some had a scan of both hands. Both datasets were manually labeled by a more than 10-year experienced radiologist as previously indicated [10] (V.F.). For the training session, images size was standardized to 256x256. A data augmentation step (*10) was performed using albumentations library [33] and included vertical and horizontal flip, arbitrary angle rotation, elastic transformation and grid distortion. The final number of images after the data augmentation process in 5600 (SDS), 12970 (LDS). Images were normalized with respect to the mean intensity and standard deviation. Each dataset ($D$) was represented by a set of 2D matrices: $D = (X_i, Y_i)_{i=1}^{N}$, where $X_i$ – is a 2D MR image of wrist and $Y_i$ – the corresponding 2D cartilage binary mask.

A cross-validation approach was used in order to estimate the achievable performance of the different CNNs on heterogeneous data. Using the GroupKFold from sklearn library [34], the dataset was divided into 5 subsets (with 20% of the total amount of 3D images in each) and used for a 5-fold cross-validation analysis. Training and testing subsets did systematically contain data from different subjects.

### 2.2. Convolutional neural networks
#### 2.2.1. Architecture and training parameters optimization

In addition to the training hyperparameters and as previously described [22], [35], number of layers (i.e., the depth of the CNN), order and presence of noise, dropout and batch normalization layers have been considered as adjustable architecture hyperparameters [22], [36]. Two variants of the U-Net architectures were used: one with an original depth, i.e. number of max-poolings equal to 4 [30], and one with a reduced depth (Truncated U-Net) i.e. with three max-poolings. Layers of batch normalization, noise and spatial dropout were added in order to improve convergence time and decrease generalization error. The order and the presence of dropout and batch normalization layers in different parts of the networks were adjusted so as to maximize the averaged 3D DSC during the testing phase. An Adam optimization algorithm was used for training with a fixed batch size (32) and cross-entropy was used as the loss function. Several hyperparameters were adjusted throughout the training process via a grid search i.e. learning rate (from $6*10^{-4}$ to $3*10^{-3}$), utilization of learning rate decay (from $50*10^{-4}$ to $5*10^{-4}$), noise level (from 0 to 0.5) and dropout probability (from 0 to 0.5).

With the aim of avoiding false positive results from regions out of the wrist joint area (an issue reported in a previous work [10]), attention layers were added in the U-net and truncated U-Net architectures. As illustrated in Figure 1, a localization layer was integrated to the skip connections as previously proposed [31]. Briefly, feature maps from encoder and decoder propagate through 1x1 size aligning convolutions and are added so that the resulting feature map values are modulated according to the feature map values of both sources. Then, feature maps are activated by a ReLU layer and normalized by a sigmoid function. The attention coefficients are then applied after a bilinear upsampling. Thus, the attention layers are intended to compute attention coefficients that are multiplied with feature maps in an element-wise manner. In that way, less informative regions can be ignored. The structure of an attention layer depicted in Figure 1 illustrates the corresponding operation in details.

All the networks were trained without using pre-optimized model's parameters using a high-performance cluster with the following characteristics: CPU: 2x Xeon Silver 4214, RAM: 128 Gb, GPU: 4x Tesla V100. CNN development was performed using Python programming language and TensorFlow and Keras open source libraries. Training was performed on all available GPUs.

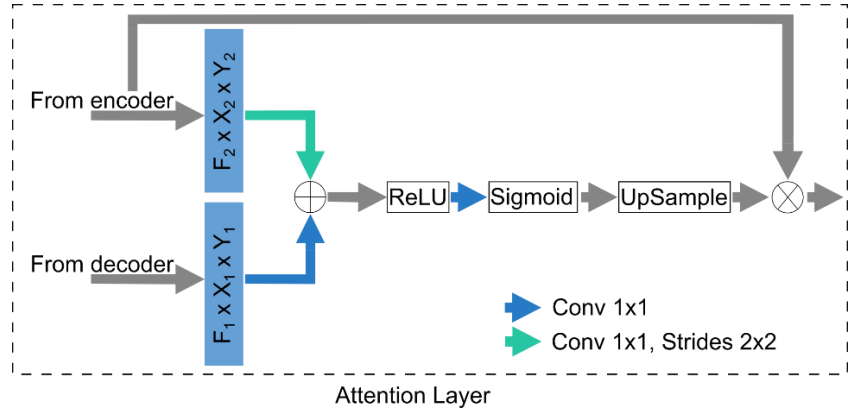

*Fig.1. Schematic structure of an attention layer. Feature maps from encoder and decoder propagate through 1x1 size aligning convolutions and are added up* added *so that the resulting feature map values are modulated according to the feature map values of both sources. After that, feature maps are activated by and normalized by a sigmoid function. Finally, after a bilinear upsampling, attention coefficients are occurred.*

## 2.3. Data analysis
### 2.3.1. Metrics and loss function

Several metrics were used in order to assess the network's performance. As a main metric we used Dice Similarity Coefficient (DSC) [25].

$$DSC = \frac{2 \cdot |\hat{Y} \cap Y|}{|\hat{Y}| + |Y|}$$

where $\hat{Y}$ - is a binary cartilage mask predicted by the CNN, and $Y$ – a ground-truth (GT) mask obtained by a human observer. DSC was calculated for each individual slice (2D DSC) and also for each of 3D images (3D DSC) in the testing datasets. In addition, an overall precision metric was computed as follows [36]:

$$Precision = \frac{TP}{TP + FP}$$

where $TP$ is the overall number of true positive pixels in the predicted masks within the whole testing set, $FP$ – the corresponding number of false positive pixels.

As previously described [10], a region-based analysis was used to assess the CNNs' performance across the wrist joint in terms of mean 2D DSC over particular wrist zones. The 2D DSC values distribution was analyzed with respect to cartilage representation. Pixels were gathered in 4 bins according to the relative amount of cartilage (#1 − 0%, #2 − 0% − 33%, #3 − 33% − 66%, #4 − 66% − 100%) and the corresponding mean DSC values were computed in these zones. In addition to DSC, cartilage volume was computed given that it is a metric of interest in joints disorders [6]. The error of cartilage volume (ΔV) measurement was computed as follows:

$$\Delta V = \frac{|V_{GT} - V_{pred}|}{V_{GT}} \cdot 100\%$$

where $V_{GT}$ is a ground truth cartilage volume computed from a manual segmentation and $V_{pred}$ is the corresponding volume computed from the automatic segmentation from a given CNN.

*2.3.2 Statistics*

Statistical analyses were performed using RSudio. One-way ANOVA analysis [37] was performed in order to assess the effect of the type of CNN on some performance characteristics (3D DSC and cartilage volume). When the effect was significant, pairwise post-hoc tests were conducted [38]. Bland-Altman [39] and Pearson correlation [40] analyses were performed in order to assess systematic bias of cartilage volume measurements and the corresponding correlation coefficients between volume computed from the manual and the automatic segmentation processes.

**3.    Results**

*3.1.    Architectures and training parameters*

Four adjusted CNN architectures were finally selected: U-Net, U-Net with attention layers (U-Net_AL), truncated U-Net (Tr_U-Net) and truncated U-Net with attention layers (Tr_U-Net_AL) (Figures A1, A2, A3, A4 in Supplementary Materials). All the variants had spatial dropout layers in the encoder and in the bottleneck (the deepest block of convolutions before upsampling). Batch normalization layers were added after every second convolution layer, except the ones in attention layers. Networks were trained with a dropout probability of 0.2 and a noise level of 0.35.

Optimal training hyperparameters were related to a given training dataset and a given network. For the training step using the SDS, a constant learning rate was optimal for U-net ($5*10^{-3}$), and for Tr_U-Net and Tr_U-Net_AL ($5*10^{-3}$). A variable learning rate (exponential decay with an initial value of $5*10^{-3}$) and a restart every 20 epochs was used as an optimal condition for U-Net_AL. For training on the *LDS*, U-Net_AL and TR_U-Net were trained with constant learning rates i.e. $5*10^{-3}$ and $3*10^{-3}$ respectively. U-Net and Tr_U-Net_AL networks were trained using an exponential learning rate decay with a restart every 20 epochs. Initial learning rates were $3*10^{-3}$ and $5*10^{-3}$ respectively.

*3.2.    CNN performance among the zones*

As illustrated in Figure 2 and as previously described, results obtained with the PB-CNN differed significantly according to the zones. DSC in zone #1 was 0.210 whereas it was significantly larger for zones #2 to 4 (DSC ranged from 0.600 to 0.730). On the contrary, based on the DSC values, performance of the U-Net-based CNNs was independent of the zones. The most significant improvement with respect to the PB-CNN was observed for zone#1 (from 0.210 to 0.811 – 0.919), a zone without cartilage, and in zone#2 (from 0.600 to 0.746 – 0.781), a zone with a small relative amount of cartilage. For zone#1, CNNs with a full architecture systematically outperformed the ones with a truncated architecture. The large 2D DSC standard deviation calculated for zone#1 can be explained by the fact that, as this zone did not contain the cartilage in the GT masks, the similarity coefficient with the predicted masks can be equal either to 0 (if the network segments false positive pixels) or 1 (if the network correctly does not segment anything).

Results of the overall 3D CNN performance are summarized in Table 1. As compared to the PB-CNN, all the U-Net-based CNNs provided a higher 3D DSC value with the U-Net_AL providing the highest value (0.81 ± 0.03). However, ANOVA test did not indicate any significant influence of the U-Net-based CNN variant on the mean 3D DSC. Duration of the CNN training with the SDS and the processing time for a single dataset (a 3D image with 88 slices) are also summarized in Table 1. All the U-Net-based networks outperformed the PB-CNN with a 20 times reduction for the processing time and a 10% reduction for the time devoted to training. The faster network was Tr_U-Net with a processing time of 2.31 s and a training time of 54h.

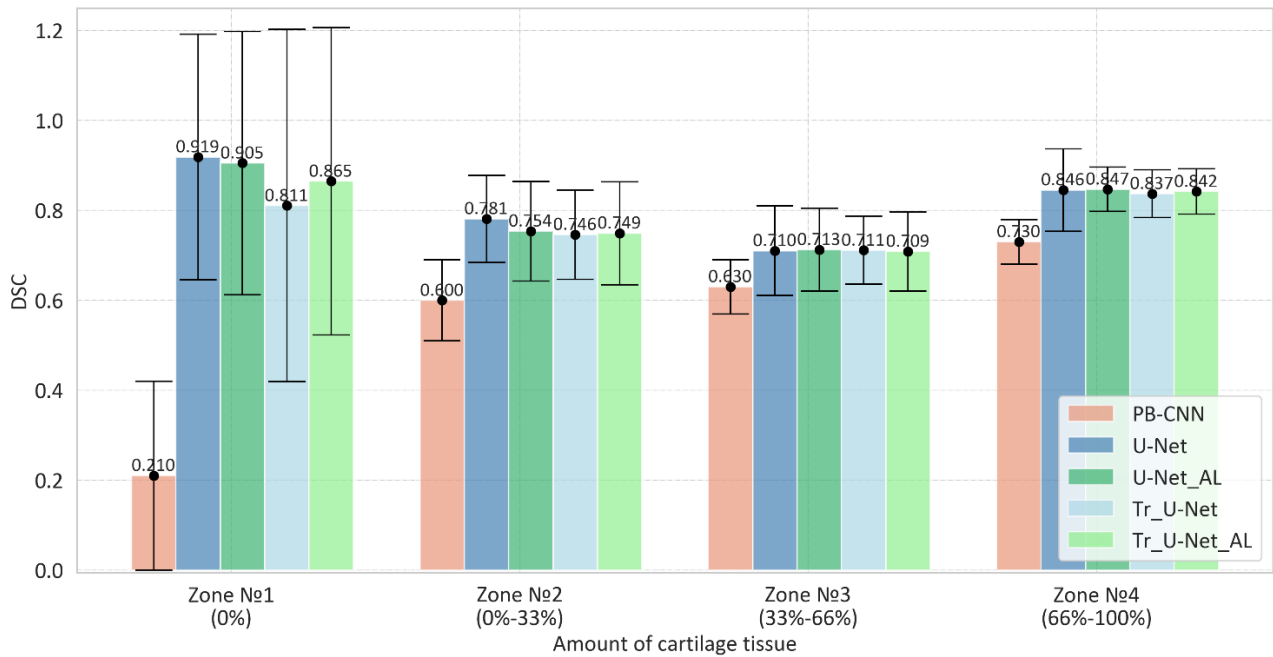

*Fig.2. 2D DSC values computed for 4 zones of wrists using 5 CNNs (trained and tested on the Small Dataset). Values are displayed as means ± SD.*

*Table.1. 3D DSC values and time performance on a Small dataset for the five CNNs*

|  | **PB-CNN** [10] | **U-Net** | **U-Net_AL** | **Tr_U-Net** | **Tr_U-Net_AL** |
|---|---|---|---|---|---|
| **3D DSC** | 0.69 ± 0.06 | 0.79 ± 0.07 | **0.81 ± 0.03** | 0.80 ± 0.03 | 0.80 ± 0.03 |
| **Processing time (s)*** | > 60 | 3.32 | 3.47 | **2.31** | 2.44 |
| **Learning time (h)** | 74.4 | 67.2 | 68.6 | 54 | **53** |

*Data were processed using a conventional computer with Nvidia GTX 1050 2 Gb, batch size -2.*

### 3.3. Cross-validation on the LDS

2D and 3D metrics were computed for all the CNNs with an extended dataset intended with the aim of assessing the CNNs performance on a more heterogeneous dataset in terms of image characteristics (voxel size, field of view, contrast, SNR, etc.). Using the cross-validation approach, we made sure to estimate the best achievable characteristics for the CNNs.

As can be seen in Figure 3, the largest median 2D DSC was acquired for U-net_AL network (see the boxplots in Figure 3 and the first line in Table 2).

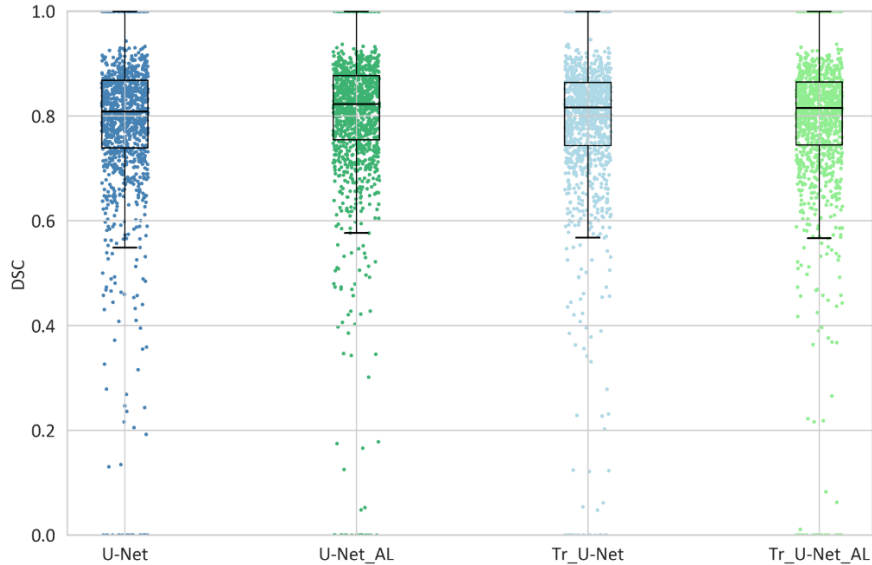

*Fig.3. Distributions of 2D DSC (merged scatter plots and boxplots), for the CNNs trained and tested on a big dataset (outliers, i.e. >Q3 + 1.5\*IQR or < Q1 – 1.5\*IQR, where IQR is an interquartile range, are excluded from the analysis).*

Volumetric DSCs for all the studied CNNs are presented in Table 2 together with the precision values. The one-way ANOVA illustrated that the 3D DSC values were influenced by the type of network (f(2)=3.88, p = 0.019). Post hoc tests indicated a significant difference between U-Net and U-Net_AL CNNs (p=0.007). U-Net_AL showed the largest DSC value i.e. 0.81 (SD = 0.037). The largest precision values were quantified for CNN with attention layers.

*Table.2. Performance metrics of the four CNNs obtained with the the Large dataset.*

| Metric | U-Net | U-Net_AL | TrU-Net | TrU-Net_AL |
|---|---|---|---|---|
| **2D DSC** | 0.808 [0.739, 0.868] | **0.822** [0.755, 0.876] | 0.817 [0.744, 0.864] | 0.815 [0.744, 0.865] |
| **3D DSC** | 0.810 [0.780, 0.822] | **0.817** [0.785, 0.838] | 0.799 [0.775, 0.828] | 0.811 [0.784, 0.823] |
| **Precision** | 0.773 | 0.780 | 0.769 | **0.781** |

*For the DSC, values are reported as median and range [25th percentile, 75th percentile]. Precision is estimated as a ratio of number of true positive pixels to the sum of true positives and false positives in the predicted masks within the whole testing set*

Figure 4 provides the results of the CNNs testing on the slices selected from different zones of a 3D image of a heathy volunteer and illustrates the architecture choice effect. Note a significant increase for 2D DSC index reduction of the number of false positive pixels in case of adding the attention layers to the U-Net network.

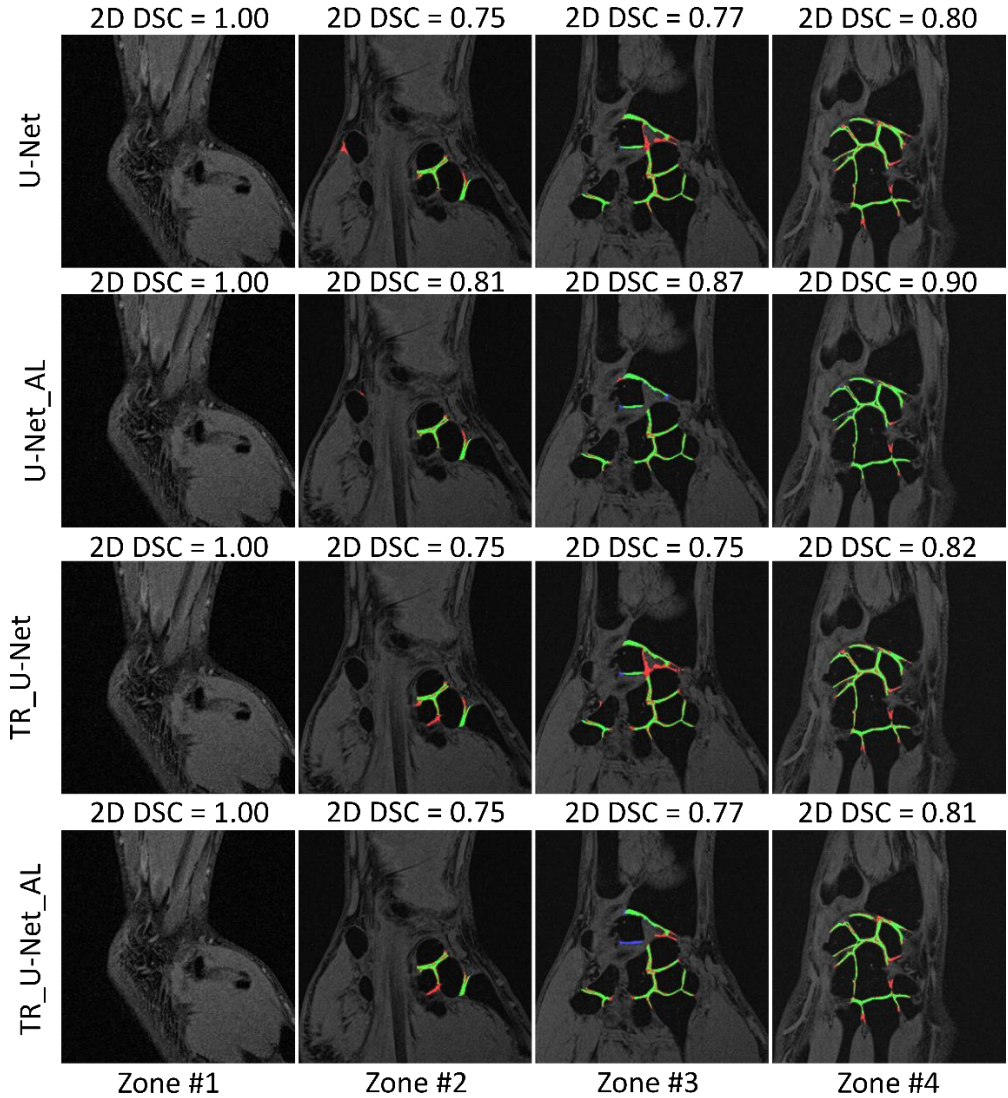

*Figure 4. Example of CNNs performance in different slices from one 3D image. Red – false positive, blue – false negative, green – true positive pixels*

### 3.4. Cartilage volume measurements

A 3D visualization of cartilage volume is provided in Figure 5. The average volume measurements computed from the manual and the automatic segmentations are summarized in Table 3. No significant CNN-effect was observed for the 3D wrist cartilage volume (f(2)=0.54, p = 0.52). As indicated in Table 3, the mean relative volume errors ranged from 17.21% (U-Net_AL) to 19.72% (U-Net). Correlations between volumes computed manually and automatically are illustrated in figure 6. The highest Pearson correlation coefficient was observed for U-Net_AL (r = 0.765) (Figure 6 a). Bland-Altman plots displayed in Figure 6 b indicated that the smallest volume difference range (±1.96σ) (green and blue solid lines) was achieved for U-Net_AL. Overall, the volume difference was independent of the volume (Figure 6 b) in all cases. For all the networks, the systematic bias ranged from 1.6 (Tr_U-Net_AL) to -67.4 (Tr_U-Net) mm$^3$.

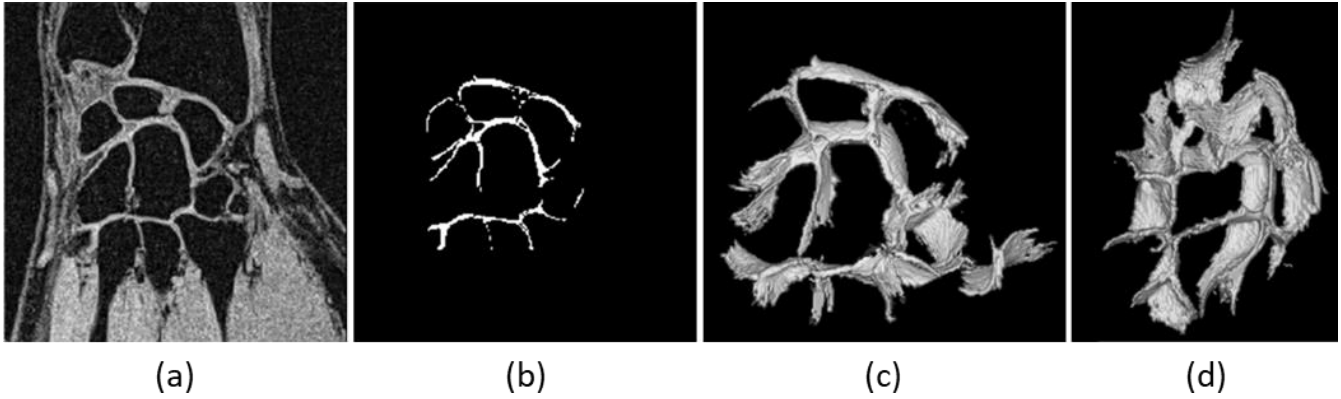

*Figure 5. (a) Medial slice from initial 3D MR image of the wrist of a healthy volunteer. (b) Corresponding cartilage mask. (c), (d) 3D image of wrist cartilage, different views.*

*Table 3. Wrist cartilage volumes and relative errors computed from the manual and the automatic segmentation*

| CNN | Volume (mm$^3$) | Mean Relative error (%) |
|---|---|---|
| **Ground Truth** | 2522 ± 704 | |
| **U-Net** | 2524 ± 683 | 19.72 |
| **U-Net_AL** | 2582 ± 624 | 17.21 |
| **TrU-Net** | 2589 ± 666 | 18.16 |
| **TrU-Net_AL** | 2520 ± 662 | 18.10 |

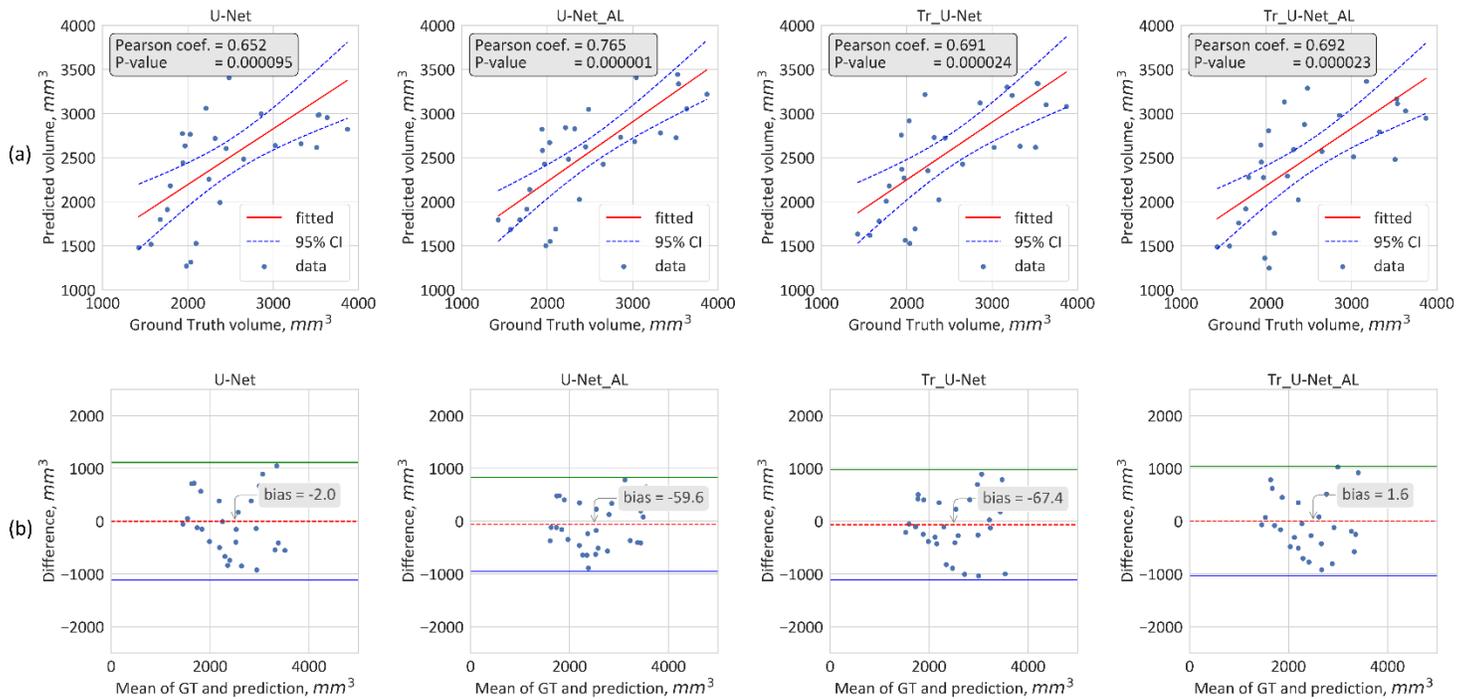

*Figure 6. (a) - Pearson correlation coefficient (r) and p-value (p) between predicted and Ground truth cartilage volume. CI – confidence interval. (b) - Bland-Altman plots for cartilage volume measurements*

In the LDS, 7 subjects were scanned repeatedly in a single session using two different coils, one of them providing higher and lower SNR [41]. The absolute difference for the cartilage volume was computed and scaled (%) to the largest value in the pair using the values from U-Net_Al and from the manual segmentation. As can be seen in Table 5, the difference was almost systematically lower for U-Net_AL with values ranging from 0.2 to 16.6 % (Table 5).

*Table 5. Cartilage volume difference from repeated measurements, %.*

| Subject | 1 | 2 | 3 | 4 | 5 | 6 | 7 | mean |
|---|---|---|---|---|---|---|---|---|
| **Manual** | 13.9 | 11.8 | 8.6 | 1.9 | 28.2 | 8.4 | 16.5 | **12.7** |
| **U-Net+AL** | 1.6 | 16.6 | 3.5 | 1.9 | 16.3 | 8.9 | 0.2 | **7.0** |

*Cartilage volume was measured from MR images segmented manually of with the U-Net-AL. Images were obtained with a home built wrist coil (higher SNR) and with a commercial extremity coil (lower SNR)*

There were three cases in the LDS when the same hand of the same subject had been scanned twice with different resolution. The difference of cartilage volume measured from the images with different resolution is summarized in Table 6. In all three cases the raise of the voxel size led to an increase of the measured volume.

*Table 6. Cartilage volume difference measured from MR images with different voxel size, %*

| Subject | 1 | 2 | 3 |
|---|---|---|---|
| **Voxel change, mm³** | 0.146x0.146x0.4 ↓ 0.195x0.195x0.4 | 0.318x0.316x0.5 ↓ 0.373x0.375x0.5 | 0.391x0.391x0.5 ↓ 0.508x0.508x0.5 |
| **Manual** | 25.87 | 5.98 | 18.28 |
| **U-Net+AL** | 0.09 | 19.82 | 12.36 |

*Cartilage volume was measured from MR images segmented manually or with the U-Net-AL. The same hand of the same subject was scanned twice with different resolution.*

## 4. Discussion and conclusion

*Image-based U-Net vs PB-CNN*

In this work, various convolutional neural networks with a U-Net-based architecture and optimized hyperparameters have been used for wrist cartilage segmentation and volumetric measurements, and the corresponding results have been compared. We hypothesized that this approach could be more efficient than the patch-based CNN approach we initially tested, and for which only the centrally located slices were properly segmented [10]. Four CNNs variants with different depth and with and without attention layers have been tested. Overall, for the same dataset, all the CNNs outperformed the patch-based CNN in terms of segmentation homogeneity and quality. Using an extended dataset, the U-Net architecture with additional attention layers provided the optimal results.

The results computed from the SDS indicate a clear benefit of image-based learning approach as compared to the previously reported patch-based strategy [10]. Patch-based approach for images segmentation has been reported as of interest given that a more comprehensive data augmentation by multiple patch views and class balancing can be performed [21], [42]. According to the present comparative analysis, these features are not adapted for an optimal 3D segmentation of wrist cartilage. It has been indicated that a PB-CNN approach could be effective as long as patches could contain enough contextual information. Our results indicate that the contextual information in each patch would not be homogenous and then not enough informative. We have already reported an issue related to generalization of a patch-based learning approach [10]. More specifically, cartilage from the lateral slices was not properly segmented likely because patches from these slices did contain a contextual information largely different than patches positioned in centrally located slices. One has to keep in mind that, in addition to that, patches with "lateral" cartilage were weakly represented in the training dataset due to the relatively smaller amount of cartilage in lateral slices. The tested U-Net-based CNNs were not only more efficient regarding the segmentation metrics but also regarding the training time. Accordingly, it has been reported that patch-based approaches were less computationally effective than fully convolution networks with an image-based learning approach [22], [43]. Our results are clearly supportive of the better performance of image-based CNNs as compared to patch-based approaches and extend this to 3D wrist cartilage segmentation.

Using an extended dataset i.e. the *LDS*, the CNNs performance was compared on the basis of a cross-validation approach. This comparative analysis was based on both geometric (DSC) and volumetric measurements. Even though the 3D DSC values were very close to each other (Table 2), the statistical analysis indicated that the median 3D DSC computed with the U-Net_AL was significantly larger. In addition, the U-Net_AL CNN provided the lowest mean volume error and the highest Pearson correlation coefficient with respect to the GT. Even though the cartilage volumes predicted by all the U-Net based CNNs were not statistically different from the other networks, the highest 3D DSC and correlation coefficient together with the smallest mean volume error illustrate that attention layers added to the U-Net CNN allowed to achieve the optimal wrist cartilage segmentation. On the contrary, the truncation of the U-Net CNN did not modify the segmentation performance.

*Attention layers*

Attention layers act as filters for feature maps that contain spatial information needed for up-convolution layers while data from the down-sampled path acts as contextual information. In other words, attention layers can minimize signal from non-contributive regions and maximize it in the regions of interest. Overall, this attention mechanism is expected to reduce false positive predictions [31]. As expected, addition of attention layers to the U-Net was linked to an increased 3D DSC and a reduced volume error while the segmentation accuracy was also raised. In the present study, the inclusion of attention mechanism allowed to improve the 3D performance although this improvement was not very critical.

*Truncation*

Truncation of the U-Net led to a reduced performance i.e. 3D DSC median value from 0.810 to 0.799. At the same time, the volume measurement error decreased from 19.72% to 18.16%. From a statistical vantage, these differences did reach the significant threshold neither for the 3D DSC nor for the cartilage volume. It should be kept in mind that the truncated U-Nets had almost a four times smaller number of parameters than the full U-Nets. Using the same batch size, the smallest processing time for a 3D image (~2.4 s) was obtained for TR_U-net+AL. In addition, the size of the network is such that the batch size could be increased with an expected further acceleration. Such an option would not be possible for the full U-Nets mainly because of the available computational resources.

*Sources of errors*

In degenerative and inflammatory diseases of joints, cartilage loss is considered as a quantitative metric of cartilage damage [1], [5], [7], [9]. Assessment of wrist cartilage volume has not been reported so far likely as a result of the complex anatomy, the cartilage size and the difficulties related to segmentation. Such a quantitative assessment would be of interest as long as the corresponding accuracy and volume error are adequate to changes reported so far in pathological situations. To the best of our knowledge, wrist cartilage volume changes in OA or RA have not been reported so far. According to studies conducted over a 10-year period in knee of osteoarthritic patients, the cartilage loss in the medial and lateral compartments would be 19.1 % and 13.8 % respective [44]. On that basis, the volume error we obtained (17.2%) with U-Net_AL should be adequate for the detection of cartilage loss in arthritis.

One has to keep in mind that our geometrical and volumetric indices were computed with respect to the ground truth i.e. the manual segmentation, which is inherently affected by errors. We previously reported [10] a 0.90 DSC value for repeated manual segmentations of wrist cross-sectional slices by the same observer. In the present study, we had the opportunity to compute cartilage volume errors from repeated MRI scanning. Interestingly, the corresponding results disclosed a lower reproducibility of manual cartilage volume measurements as compared the U-Net_AL based quantification. This may indicate that the network that has been trained on a heterogeneous data, both in terms of image quality and manual labelling quality, provides a more reproducible result than the human observer, that indicates an optimal generalization ability of the CNN. Another important issue that could affect cartilage volume quantification is related to MR image resolution. On the basis of microcomputed tomography, it has been reported that cartilage thickness in a healthy wrist joint varied according to location from 0.6 to 1 mm [45]. In our study, the voxel size was within this range i.e. between 0.146x0.146x0.4 mm$^3$ and 0.508x0.508x0.5 mm$^3$. Although the sample size was small for averaging and generalizing the results, the results obviously show that the MRI-based wrist cartilage volume is strongly affected by the image resolution. These findings indicated that even though the manual labels are considered as the ground truth in our research, they cannot be considered as a "golden standard". For this reason, a more precise validation of wrist cartilage volume in future should rely on independent non-MRI-based measurements, for example, direct measurement of cartilage volume in cadaver joins [46].

*Limitations*

We have to acknowledge that the datasets used in the present study were mainly composed of healthy volunteers with a few patients only. On that basis, for a potential clinical application, one could have to fine-tune the network reported in the present study. As a follow up of the present study, it could be of interest to perform multiclass segmentation for cartilage of each carpal bone similarly to what has been done for knee cartilage [6], [44], [47]. Atlas-based [3], graph-based [48], multiclass CNN-based [49] tools could be the useful tools for this multi-class segmentation task.

## 4. Conclusion

U-Net convolution neural networks provided a significantly higher segmentation homogeneity within a 3D wrist VIBE image than a previously proposed PB-CNN. U-Net with additional attention layers provided the best segmentation quality. In order to be used in clinical conditions, our network can be fine-tuned on a dataset representing a group of specific patients. The error of cartilage volume measurement should be assessed independently using a non-MRI method in order to estimate the accuracy of the method. These additional studies will strengthen the obtained results and potentially further improve the segmentation quality.

## 5. Acknowledgements

This work was supported by the Ministry of Education and Science of the Russian Federation (075-15-2021-592). Part of the work related to magnetic resonance imaging at 1.5T was supported by a grant for scientific school НШ-2359.2022.4

**Supplementary materials**

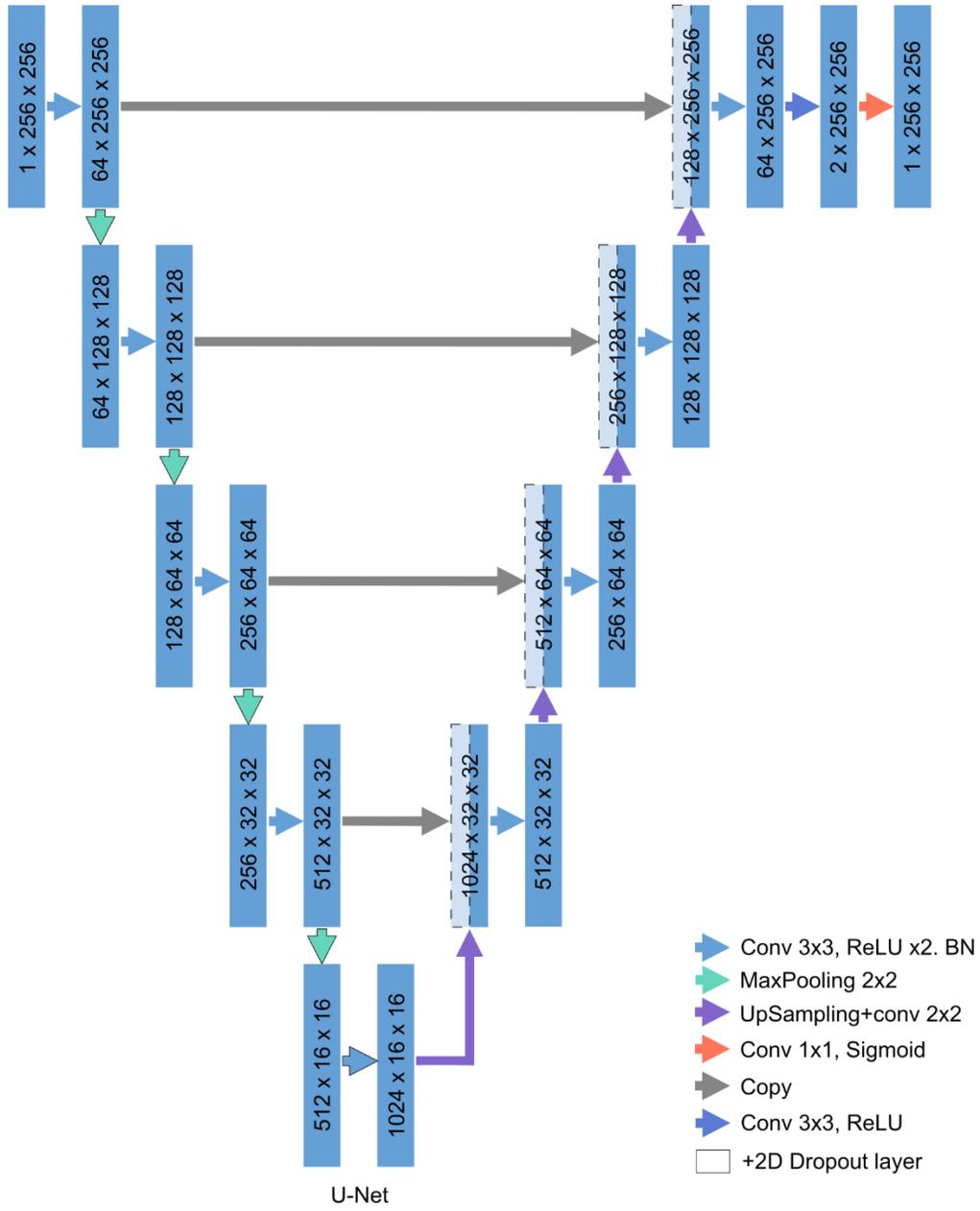

*Figure A1. Architecture of the U-Net convolutional neural network*

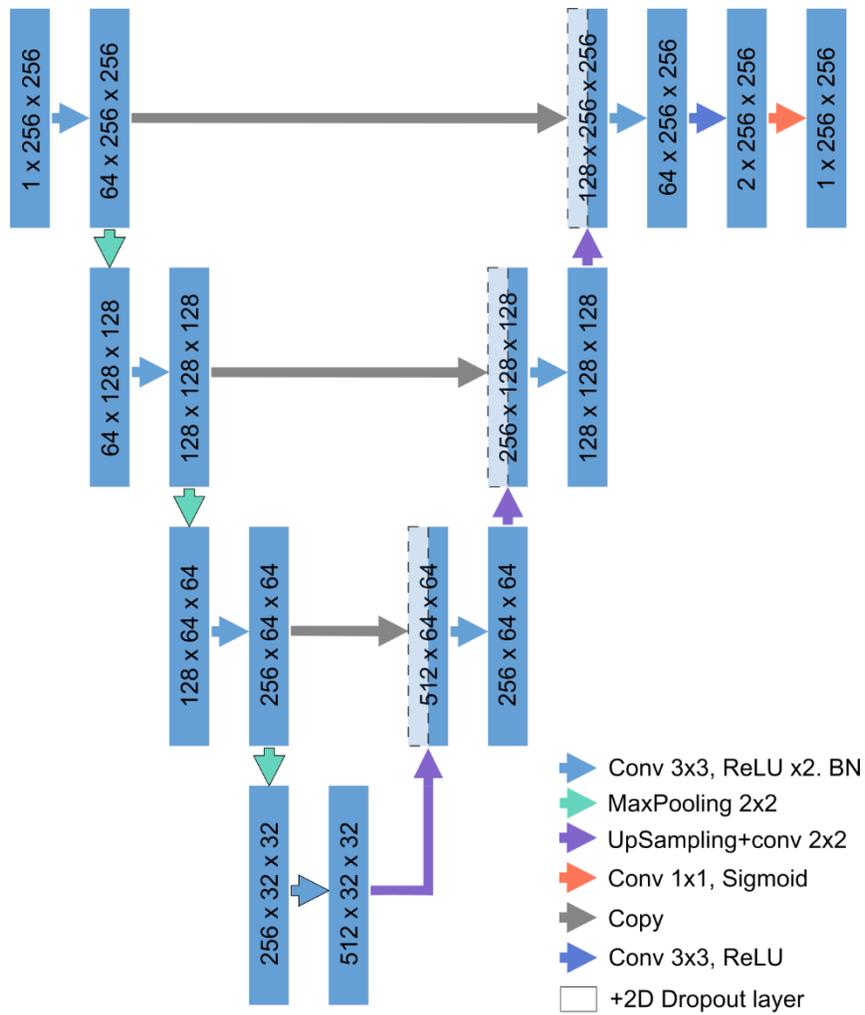

*Figure A2. Architecture of the truncated U-Net convolutional neural network.*

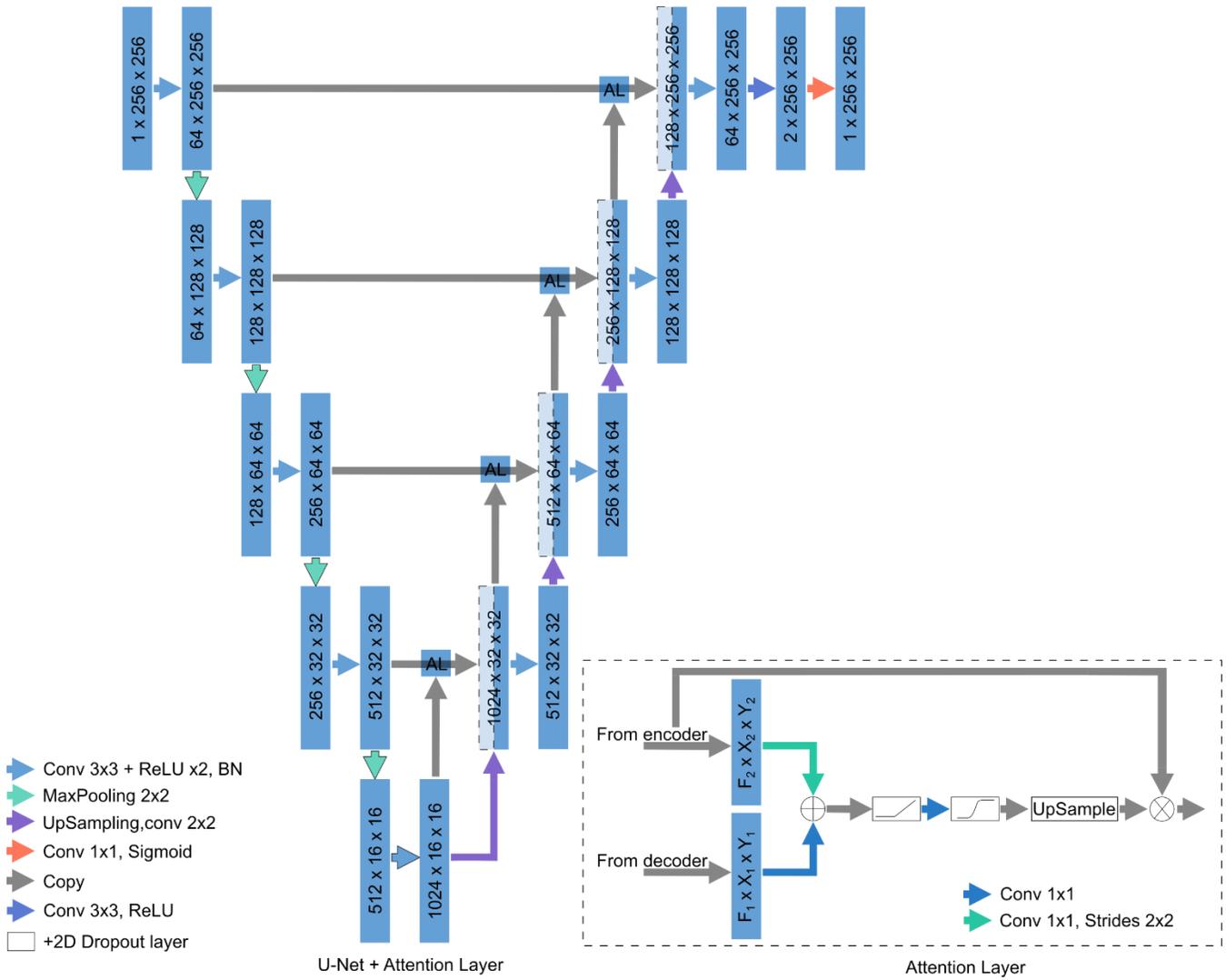

*Figure A3. Architecture of the U-Net convolutional neural network enriched with attention layers (AL). The attention layers were built into skip connections allowing to use information from the encoder path and to filter it eventually emphasizing important regions by combining low- and high-level information from both paths.*

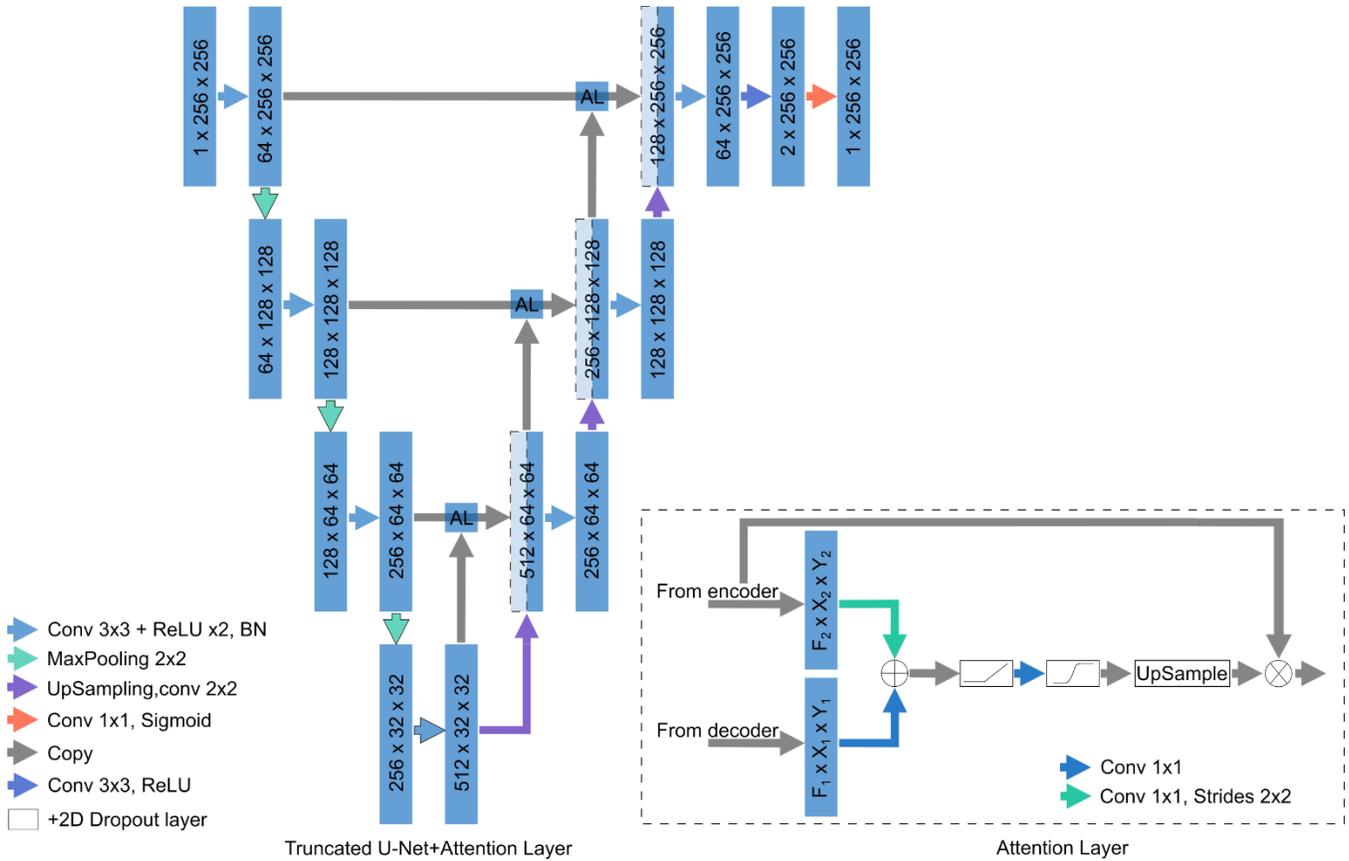

*Figure A4. Architecture of the truncated U-Net convolutional neural network enriched with attention layers (AL). The attention layers were built into skip connections allowing to use information from the encoder path and to filter it eventually emphasizing important regions by combining low- and high-level information from both paths.*

*MR imaging details*

Images from the SDS were obtained on a 1.5T Magnetom Espree system (Siemens) with a 3D VIBE sequence (TR/TE = 18.6/7.3 ms, flip angle = 10°, FOV = 97x120 mm2, matrix size = 260x320, voxel size = 0.37x0.37x0.5 mm3). Scanning was performed using either a home-built volumetric wireless coil [41], or a transmit receive CP Extremity coil (Siemens). This dataset has been previously described [10]. It was halved in two subsets, one for training (50%) and one for testing (50%). The validation dataset used during the training session was set to 10% of the training dataset.

Additional images from the LDS were acquired at 1.5T (Siemens Magnetom Espree) and at 3 T (Siemens Magnetom TrioTim) using a 3D VIBE pulse sequence and either a volumetric wireless coil [41] or a CP extremity coil). The parameters were in the following ranges: FOV varied between 75x75 mm2 and 130x130 mm2, TR - 10 ms - 18.6 ms, TE - 3.38 ms - 7.6 ms, voxel size - 0.146x0.146x0.4 mm3 - 0.508x0.508x0.5 mm3, matrix size - 256x256 - 260x320. The flip angle was constant (10°) in all variants of the pulse sequence.